%% file: sample-sigconf.tex
\documentclass[sigconf]{acmart}

\usepackage[utf8]{inputenc}

\usepackage{balance} 
\usepackage{booktabs} 

\copyrightyear{2018}
\acmYear{2018}
\setcopyright{rightsretained}
\acmConference[DTUC '18]{Digital Tools \& Uses Congress}{pp 1--4, October 3--5, 2018}{Paris, France}
\acmBooktitle{Digital Tools \& Uses Congress (DTUC '18), pp 1--4, October 3--5, 2018, Paris, France}\acmDOI{10.1145/3240117.3240118}
\acmISBN{978-1-4503-6451-5/18/10}

\begin{document}
\title{Predicting Personalized Academic and Career Roads: First Steps Toward a Multi-Uses Recommender System}


\author{Alexandre Nadjem}
\orcid{1234-5678-9012}
\affiliation{%
  \institution{Avignon Universit\'e, France}
  \streetaddress{Adresse 1}
  \city{}
  \state{}
  \postcode{1234-567}
}
\email{alexandre@humanroads.fr}

\author{Juan-Manuel Torres-Moreno}
\orcid{1234-5678-9012}
\affiliation{%
  \institution{Avignon Universit\'e, France}
  \streetaddress{Adresse 1}
  \city{}
  \state{}
  \postcode{1234-567}
}
\email{juan-manuel.torres@univ-avignon.fr}

\author{Marc El-Beze}
\orcid{1234-5678-9012}
\affiliation{%
  \institution{Avignon Universit\'e, France}
  \streetaddress{Adresse 1}
  \city{}
  \state{}
  \postcode{1234-567}
}
\email{marc.elbeze@univ-avignon.fr}

\author{Guillaume Marrel}
\orcid{1234-5678-9012}
\affiliation{%
  \institution{Avignon Universit\'e, France}
  \streetaddress{Adresse 1}
  \city{}
  \state{}
  \postcode{1234-567}
}
\email{guillaume.marrel@univ-avignon.fr}

\author{Beno\^it Bonte}
\orcid{1234-5678-9012}
\affiliation{%
  \institution{Millionroads, Avignon, France}
  \streetaddress{Adresse 1}
  \city{}
  \state{}
  \postcode{1234-567}
}
\email{benoit@humanroads.com}
\renewcommand{\shortauthors}{A. Nadjem et al.}

\begin{abstract}
Nobody knows what one's do in the future and everyone will have had a different answer to the question : how do you see yourself in five years after your current job/diploma? In this paper we introduce concepts, large categories of fields of studies or job domains in order to represent the vision of the future of the user's trajectory. 
Then, we show how they can influence the prediction when proposing him a set of next steps to take.
\end{abstract}

%
%

\ccsdesc[500]{Human-centered computing~Social networks}
\ccsdesc[300]{Human-centered computing}
\ccsdesc[300]{Applied computing~Document search}

\keywords{Academic roads, Career roads, Recommender systems, Statistical analysis}

\maketitle

\input{samplebody-conf}

\balance
\bibliographystyle{ACM-Reference-Format}
\bibliography{sample-bibliography}

\end{document}

%% file: samplebody-conf.tex
\section{Introduction}

Identifying the next possible step in a career involves many different factors. They can be hidden, like personal reasons, or specific to a time period, often to a more general context, like the reputation of a company. Others can  also be explicit like the skills and the past job in a \textsl{résumé} \cite{KESSLER}. Different strategies have been studied to predict which will be the next job or company.
For instance, it is possible to find hidden mechanisms in a career evolution after investigating a specific field or job. 
\cite{James2018} chose to focus on the career evolution of researchers and improved the prediction of the next workplace of a researcher by excluding the laboratory a researcher had no contact or never worked with.
\cite{Li2017} have studied a set of real LinkedIn\footnote{\url{http://www.linkedin.com}} data to build a next career prediction system. By crossing multiple information not only about the user's past but also about the company, they improve the precision on not only the next job but also the next location.

We cannot compare with most works focused on predicting the next step in a career since they use the assumption it will be in the same field as the last one. At the beginning of their studies, few people have a clear objective and actually pursue it. Some follow a standard path, others hear about an opportunity and go for something they would never have expected. By increasing the number of proposed career paths, one can find new recommendations that would motivate this person. This is why we think it is important to display as many choices as possible avoiding to represent the evolution of a career with linear or standard modeling. We firmly believe that the user must not feel driven to a specific place but a set of opportunities where to look for.

\section{Problem definition}

A lot of information is hidden when looking only at the \textsl{résumé} but there might be hints in the past description useful to find clues of what has influenced some change. Maybe someone was learning music
in its free time and after many years this person decided to go back to this activity.
At first, we should define what a reorientation is. Reorientation may be understood in many different ways\cite{Negroni05}.
So far, we do not want to choose one of them.
Nevertheless, from the moment that we agree to take into account this phenomenon, many tracks of research are open at 3 levels: data analysis, models for prediction, multi-uses of recommender systems (RS). 
From an \textsl{analysis point of view}, can we find on each trajectory a clue allowing the hypothesis that a reorientation occurred? Is it possible to find classes of reorientation\footnote{Some reorientations could be slow and prepared, like someone starting a new diploma or a vocational education, or could happen suddenly without giving any warning, like changing from trader to baker.} using or not a preset of categories? 
From a \textsl{prediction point of view}, there are mainly two problems. The first one is how to differentiate trajectories with reorientations from ones without. 
When there is no reorientation, recommending the continuity seems obvious. On the contrary, assuming that a reorientation is sure, the system will have to choose between many possible new activities. As done in Information Retrieval with the Relevance Feedback principle \cite{KarenSparck,CABRERA}, how can we filter all these choices and put the user in the loop?
From a \textsl{RS point of view}, are we able to explain to the user why the system gave these results by displaying the different hints used to make one or another proposition?  
In the experiments reported in Section 5, we will focus on the prediction level with the purpose to reuse it soon in some RS functionalities. 

\section{Data representation}

\subsection{Data's Fragility}

An on-line \textsl{résumé} is composed of declarative sentences. Without a more objective data source, the RS is still dependent of the subjectivity of the autobiographical writing and the goal of creating a self-introduction \cite{Lejeune,AGGERMAN}.

Any biographical work produces an illusion \cite{Bourdieu}, but we accept the risk of categorizing these massive declarative data.

\subsection{Recoding for normalization}

Two graduated students who studied in the same school and got the same diploma will not present it the same way in their \textsl{résumés}. One might use the complete name, the other an abbreviation (Bachelor of Business Administration/B.B.A). We need to address this diversity and regroup similar steps under the same entities or categories by normalizing incoming data. We have used the nomenclature from the ONISEP\footnote{\url{http://www.onisep.fr/}} as a model for standardizing the steps composing a trajectory and for categorizing the profession\cite{Desrosieres,Amosse}. The International Standard Classification of Occupations is a tool for organizing jobs into a clearly defined set of groups according to the tasks and duties undertaken in the job. It is intended both for statistical uses and for client oriented uses\footnote{\url{http://www.ilo.org/public/english/bureau/stat/isco/isco08/index.htm}}.

\subsection{Dataset}

In  this paper, we are using data coming from Viadeo\footnote{\url{http://www.viadeo.com}}, a professional social network. We also take advantage from previous works done by HumanRoads\footnote{\url{http://www.humanroads.com}} covering the extraction of a list of French diplomas, job titles and their translations.

After analyzing a \textsl{résumé}, either it has been written on paper or on a computer, it is possible to extract different categories of information. 

\begin{itemize}
\item User: A user is unique and can be represented by a name, an email or an id. For the prediction, they can be anonymous, because we focus on what the users have done and not on their real identity.

\item Steps: these data contain the highest amount of information. Each step is composed of title \textsl{T}, start and  end date, location, additional information like detailed description of tasks and knowledge acquired.

\item Skills: it is a description of what has been learned over the years.  A user can underline some skills and fields. He can give an appreciation on himself in a particular field or get it from somebody else.

\end{itemize}
The dataset is composed of 9383 users and a total of 65403 steps (i.e. an average of 6.9 steps per user).

\subsection{Fields and Concepts}

Fields are information/tags such as “internet” or “wind-power” that categorize the current step. Concepts (\textsl{C}) are even larger categories regrouping a maximum of fields while being distinct enough from each other, like “computer science” (CS) or “environment \& energy”. The concepts are large enough to simulate a fuzzy vision of the future.  When someone asks for help, he often has a vague idea of what he wants to do next. We use the concepts as hints to simulate someone explaining ``I want to work/study somewhere related to environment or energy”. We have 17 Concepts for diplomas and 47 for jobs.

\subsection{Approaches for profile modeling}\label{pmodeling}
A profile can be modeled as a succession of steps. 
As shown in the example given in Table \ref{tab:freq}, the first 3 steps represent diplomas. Step 4 represents the first job done after the studies. Each step is composed of keywords (like CS), which help to classify the steps under the corresponding concepts.

\begin{table}[h!]
  \caption{Steps of a user}
  \label{tab:freq}
  \begin{tabular}{l|l|l|l|l}
    \toprule
    \bf  & \bf Step 1 &\bf  Step 2 &\bf  Step 3 &\bf  Step 4\\
    \midrule
 \hline
    \textsl{T} & HighSchool & Bachelor   & Master   & Software  \\
     & degree & in CS   & in CS  & Consultant  \\ 
\hline    \hline
    $C_1$ & \textit{\textbf{Math \&}}   & \textit{\textbf{CS \&}}  & CS \&  & \textit{\textbf{CS \&}} \\
    & \textit{\textbf{Science}} & \textit{\textbf{Internet}}   & Internet   &  \textit{\textbf{Internet}} \\ \hline
    $C_2$ & & & & \textit{\textbf{Management}} \\
    & & & & \textit{\textbf{\& consulting}}\\ 
\hline \hline
	 & \bf  Past & \bf Present &      Future & \bf User intention \\
  \bottomrule
\end{tabular}
\end{table}

Baseline: for everybody, the RS will always propose, at each step, the same list ordered by decreasing frequencies. In other words, the context is completely ignored here.
Looking for a better prediction, we use one of three key elements at a time to improve the recommendations. The first job after the current step, (this first job could be the next step but also a later one), the highest diploma obtained up to date and the concept of the previous step. 

To simulate user interactions with the system, we need to have a ``future'' step. We extract fields and concepts out of the next step as a feedback from the user for the prediction. Since step 4 is the last one, we will not try to predict it. We also need more information about the past of the user, thus we will also not predict step 1. We removed all the profiles with less than 3 steps from the dataset. If a step is not classified, we also remove it.
Back to Table \ref{tab:freq}, note that,to predict step 2, we only need information highlighted in italics and bold.

\section{Algorithm: Next field prediction}
\begin{figure}
\includegraphics[height=4.5cm]{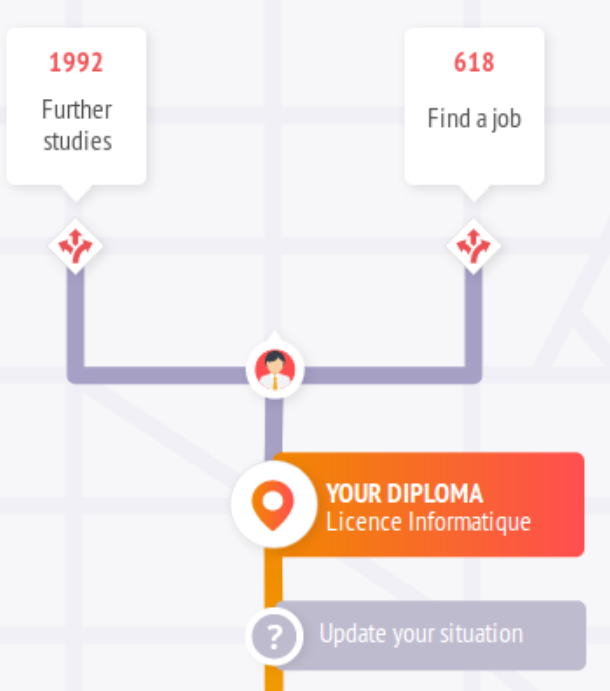}
\caption{1\textsuperscript{st} option given to the user after a bachelor in CS.}
\label{fig:1}
\end{figure}

The HumanRoads tool developed for visualization gives us a good basis for modeling interaction with the user. The following examples showcase a user whose current step is bachelor in CS. 
He can access the path shown in Figure 
\ref{fig:1}.

\begin{figure}[h!]
 \setlength{\belowcaptionskip}{-15pt}
\includegraphics[height=5.5cm,width=8.5cm]{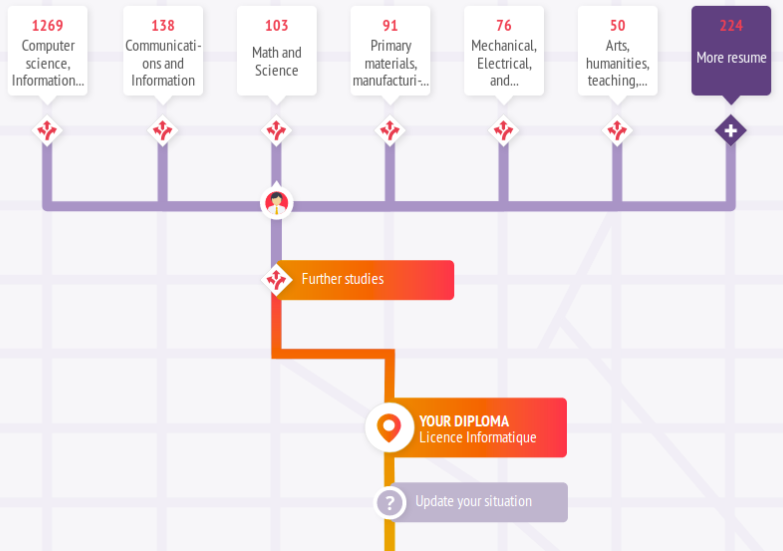}
\caption{2\textsuperscript{nd} option given to the user.}
\label{fig:2}
\end{figure}

\noindent If a user chooses “Further studies”, we can ask him 2 questions leading to different options: the first one would be, do you already have a goal? If the user already knows what he wants, he is asking for additional information. Some options include the shortest way to achieve his goal, the most commonly chosen studies or intermediate jobs.
But if his answer is ``No'', we need more precision to propose an orientation. Maybe he has a vague idea of what he wants to study. We propose the first six most commonly chosen concepts (see Figure 
\ref{fig:2}) and a list of the remaining concepts in “more \textsl{résumé}” (see Table \ref{tab:moreresume}). Since it is not possible to massively involve users during the experimentation, we have simulated these interactions by looking at concepts of the next steps. 

\newpage
\begin{table}
\caption{Other concepts in more \textsl{résumé}.}
  \label{tab:moreresume}
  \begin{tabular}{l|c}
    \toprule
    \bf Concepts & \bf Frequency\\
    \midrule
    Army \& Security   & 42 \\
    Business, Sales \& Marketing & 41 \\ 
    ... & ... \\ 
    Agriculture, fishing  & 10 \\
  \bottomrule
\end{tabular}
\end{table} 


\subsection{A simple model}
At this stage, the model (\ref{eq:model}) is purposely kept very simple in order to favor the explanatory dimension.
Given a set of observations (namely the concepts $C$ on which we can rely), the hypotheses $H$ are sorted for the prediction through a probabilistic decision-making approach $P(H|C)$. 
The frequencies $F(H,C)$ are sorted by the $S$ function, in a decreasing order, guaranteeing  the optimality  without any approximation thanks to the Bayes rules (see eq. \ref{eq:model}).
\vspace{-1pt}
\begin{equation}
   \tilde{S}(H|C) = \frac{S_H P(H,C)}{P(C)}=S_H P(H,C)=S_H F(H,C)
   \label{eq:model}
\end{equation}
\subsection{Evaluation criterion}
For evaluation purpose, we have adapted the well-known Reciprocal Rank measure (MRR) \cite{Voorhees1999}. If the user goes for the first choice, we score 1, the second $\frac{1}{2}$ and so on. 
Since, for each bucket of 6 propositions, they are displayed more or less on the same level,
we apply a "fudge factor" $\alpha$ softening the difference of penalties between two consecutive ranks $r$, in the same pack:
$s(r)$  = $(1/r)^{\alpha}$ (with $\alpha$ empirically set to 0.2).
If none of the results are correct, the answer is hidden in “more \textsl{résumé}”. Since it requires a new action to develop a new list, we divide the following score by 2. Every 6 propositions, we divide again by 2 the scores, because it requires an additional effort for the user to find a fitting proposal.

\section{Experiments}
After removing steps mentioned in section \ref{pmodeling}, it remains $7.500$ users ($N$), $17.500$ diploma steps ($N_d$),  $24.000$ job steps ($N_j$).
In order to respect the principle of a non-biased evaluation, we opted for a cross-validation process. Before predicting a step, this one is temporarily removed from the dataset and the $Nj-1$ remaining steps are used as a training set.

\subsection{Results}
\begin{figure}
\includegraphics[height=4.8cm]{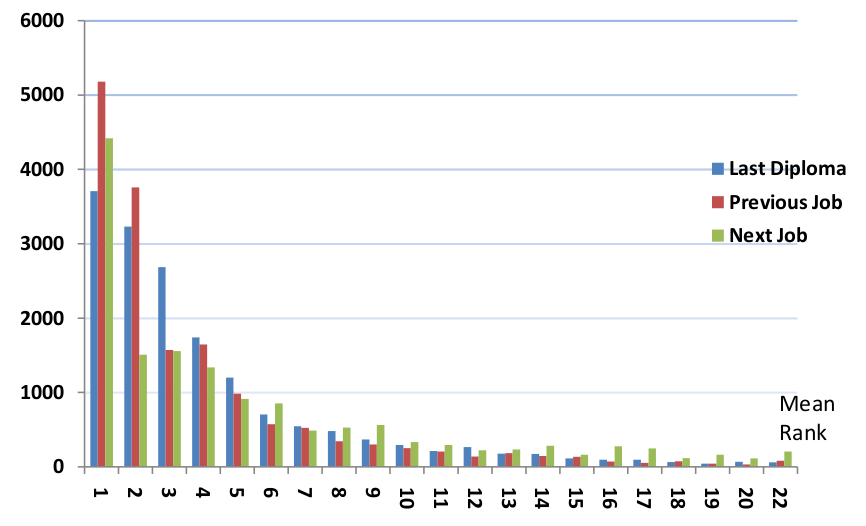}
\caption{MR for Concept prediction of current Job.}
\label{fig:concept}
\end{figure}

Figure \ref{fig:concept} shows the Mean Rank (MR) for Concept prediction of the current job relying on 3 possible Concepts : Previous job, Last Diploma or Next job. The histogram shows the number of steps in each interval [r,r+1]. 
Clearly, relying on the previous job gives the best density in the lowest ranks.

In Table \ref{tab:MRRdiploma}, the MRR criterion (3\textsuperscript{rd} column) allows to compare 4 methods applied to predict concepts for the current diploma (from 0.73 to 0.75). The confidence interval (CI) is given for the MRR in the 4\textsuperscript{th} column.
Finally, Table \ref{tab:MRR} shows
the MRR for the jobs description.
Using this criterion, 
we compare 4 methods to predict concepts for the current job (from 0.73 to quite 0.8). Both for diplomas and jobs, relying on any information site outperforms the Baseline.
\begin{table}[h]
\caption{MR and MRR criterion comparison of four methods to predict concepts for the current diploma.}
  \label{tab:MRRdiploma}
  \begin{tabular}{l|l|l|c}
    \toprule
    \bf Method & \bf MR & \bf MRR & \bf CI\\
    \midrule
    Baseline   & 4.7 & 0.699 & [0.692, 0.706] \\
    First Job    & \bf 4.0 & \textbf{0.750} & \textbf{[0.743, 0.756]} \\ 
    Previous Diploma & \bf 4.0 & 0.733 & [0.727, 0.740] \\ 
    Higher Level Diploma & \bf 4.0 & 0.740 & [0.734, 0.747] \\
  \bottomrule
\end{tabular}
\end{table}

\begin{table}[h]
\caption{MR and MRR criteria comparison of four methods to predict concepts for the current job.}
  \label{tab:MRR}
  \begin{tabular}{l|l|l|c}
    \toprule
    \bf Method & \bf MR & \bf MRR & \bf CI\\
    \midrule
    Baseline  & 5.1 & 0.730 & [0.724, 0.737] \\
    Last Diploma    & 4.3  & 0.763 & [0.756, 0.769] \\ 
    Previous Job    & 3.8 & 0.798 & [0.791, 0.804] \\ 
    Next Job    & \textbf{3.7} & \textbf{0.8} & \textbf{[0.797, 0.811]} \\
  \bottomrule
\end{tabular}
\end{table}

\subsection{Discussion}
Using the concept of the first job after the current diploma gives the highest prediction. This suggests that the intention after a diploma has a higher impact in the choice of a career. This also emphasizes the need to interact with the user and include him in the decision-making. 
%

If the current step is a job, the concept of the previous job and the next job results are close. The lower bound using the previous job (0.791) is higher than the upper bound for a diploma (0.740) showing a higher continuity between 2 consecutive steps in a professional career than in an academic one.
Once again, using the user's intention gives the best results. The MR decreases to 3.7 in this case. 

With an interval ranging from 0.797 to 0.811, the MRR has been upgraded up to 0.8.
A part of the remaining 20\% may be due to the simplicity of our methods; the rest must come from reorientations and  their unpredictability.
In section 2, we choose not to define explicitly what a reorientation is. Now, we can consider it as the set of steps the system has not correctly predicted, (those included in the least probable results).
This affirmation has been confirmed by many samples we found when analyzing logs of the decision process. For instance, someone working for years in hotels (tourism domain), suddenly and singularly, move to health care which is the 13\textsuperscript{th} most popular hypothesis over 47. How could this proposal be predicted at a better rank, close to the MR 3.7? 

\section{Conclusions and short term perspectives}

In this paper, we have studied how to predict the next step of academic or career roads without taking into account a possible reorientation. We have used concepts induced from the future in order to simulate the fuzzy vision of the user intentions.  

We are aware that using large categories, even if they are distinct enough, has some impact on the results. It will be straightforward to use a higher level of granularity such as the fields. 
Now, the system recommends for the current step an ordered list of concepts at a rough level. In order to cope with the non linear distribution, we could model such a long tail, regrouping the weak frequencies. Beyond this, we are currently working on the opposite angle:  
a RS designed to rank and find the best profiles matching a job concept. This way, it will be a new opportunity to search how to include the principle of mobility in the model.




\begin{acks}

  The authors would like to thank the anonymous referees for their valuable comments and helpful suggestions. The work is supported by the \grantsponsor{GS501100001809}{ANRT France}  
~\grantnum{2017/0400}{ Grant No.:2017/0400}.

\end{acks}

%% file: sample-sigconf.bbl

\begin{thebibliography}{12}


\ifx \showCODEN    \undefined \def \showCODEN     #1{\unskip}     \fi
\ifx \showDOI      \undefined \def \showDOI       #1{#1}\fi
\ifx \showISBNx    \undefined \def \showISBNx     #1{\unskip}     \fi
\ifx \showISBNxiii \undefined \def \showISBNxiii  #1{\unskip}     \fi
\ifx \showISSN     \undefined \def \showISSN      #1{\unskip}     \fi
\ifx \showLCCN     \undefined \def \showLCCN      #1{\unskip}     \fi
\ifx \shownote     \undefined \def \shownote      #1{#1}          \fi
\ifx \showarticletitle \undefined \def \showarticletitle #1{#1}   \fi
\ifx \showURL      \undefined \def \showURL       {\relax}        \fi
\providecommand\bibfield[2]{#2}
\providecommand\bibinfo[2]{#2}
\providecommand\natexlab[1]{#1}
\providecommand\showeprint[2][]{arXiv:#2}

\bibitem[\protect\citeauthoryear{Bourdieu}{Bourdieu}{1986}]%
        {Bourdieu}
\bibfield{author}{\bibinfo{person}{Pierre Bourdieu}.}
  \bibinfo{year}{1986}\natexlab{}.
\newblock \showarticletitle{L'illusion biographique}.
\newblock In \bibinfo{booktitle}{\emph{Actes de la Recherche en Sciences
  Sociales}}. \bibinfo{publisher}{Persée}, \bibinfo{address}{Lyon, France}.
\newblock


\bibitem[\protect\citeauthoryear{Cabrera-Diego, El-Bèze, Torres-Moreno, and
  Durette}{Cabrera-Diego et~al\mbox{.}}{2019}]%
        {CABRERA}
\bibfield{author}{\bibinfo{person}{Luis~Adrián Cabrera-Diego},
  \bibinfo{person}{Marc El-Bèze}, \bibinfo{person}{Juan-Manuel Torres-Moreno},
  {and} \bibinfo{person}{Barthélémy Durette}.}
  \bibinfo{year}{2019}\natexlab{}.
\newblock \showarticletitle{Ranking résumés automatically using only
  résumés: A method free of job offers}.
\newblock \bibinfo{journal}{\emph{Expert Systems with Applications}}
  \bibinfo{volume}{123} (\bibinfo{year}{2019}), \bibinfo{pages}{91--107}.
\newblock


\bibitem[\protect\citeauthoryear{Desrosières and Thévenot}{Desrosières and
  Thévenot}{1988}]%
        {Desrosieres}
\bibfield{author}{\bibinfo{person}{Desrosières} {and}
  \bibinfo{person}{Thévenot}.} \bibinfo{year}{1988}\natexlab{}.
\newblock \bibinfo{booktitle}{\emph{Les catégories socioprofessionnelles}}.
\newblock


\bibitem[\protect\citeauthoryear{J.A., R.M., and M.J.}{J.A.
  et~al\mbox{.}}{2011}]%
        {AGGERMAN}
\bibfield{author}{\bibinfo{person}{Baggerman J.A.}, \bibinfo{person}{Dekker
  R.M.}, {and} \bibinfo{person}{Mascuch M.J.}} \bibinfo{year}{2011}\natexlab{}.
\newblock \bibinfo{booktitle}{\emph{Controlling Time and Shaping the Self:
  Developments in Auto­biographical Writing since the 16th Century}}.
  \bibinfo{series}{Egodocuments and History Series}, Vol.~\bibinfo{volume}{3}.
\newblock \bibinfo{publisher}{Brill}.
\newblock


\bibitem[\protect\citeauthoryear{James, Pappalardo, Sirbu, and Simini}{James
  et~al\mbox{.}}{2018}]%
        {James2018}
\bibfield{author}{\bibinfo{person}{Charlotte James}, \bibinfo{person}{Luca
  Pappalardo}, \bibinfo{person}{Alina Sirbu}, {and} \bibinfo{person}{Filippo
  Simini}.} \bibinfo{year}{2018}\natexlab{}.
\newblock \showarticletitle{Prediction of next career moves from scientific
  profiles}.
\newblock \bibinfo{journal}{\emph{ArXiv [stat.AP]}} \bibinfo{volume}{1802},
  \bibinfo{number}{04830} (\bibinfo{year}{2018}), \bibinfo{pages}{36--44}.
\newblock
\urldef\tempurl%
\url{https://arxiv.org/abs/1802.04830}
\showURL{%
\tempurl}


\bibitem[\protect\citeauthoryear{Kessler, Béchet, Roche, Torres-Moreno, and
  El-Bèze}{Kessler et~al\mbox{.}}{2012}]%
        {KESSLER}
\bibfield{author}{\bibinfo{person}{Rémy Kessler}, \bibinfo{person}{Nicolas
  Béchet}, \bibinfo{person}{Mathieu Roche}, \bibinfo{person}{Juan-Manuel
  Torres-Moreno}, {and} \bibinfo{person}{Marc El-Bèze}.}
  \bibinfo{year}{2012}\natexlab{}.
\newblock \showarticletitle{A hybrid approach to managing job offers and
  candidates}.
\newblock \bibinfo{journal}{\emph{Information Processing \& Management}}
  \bibinfo{volume}{48}, \bibinfo{number}{6} (\bibinfo{year}{2012}),
  \bibinfo{pages}{1124 -- 1135}.
\newblock


\bibitem[\protect\citeauthoryear{Lejeune}{Lejeune}{1989}]%
        {Lejeune}
\bibfield{author}{\bibinfo{person}{Philippe Lejeune}.}
  \bibinfo{year}{1989}\natexlab{}.
\newblock \bibinfo{booktitle}{\emph{On Autobiography}}.
  \bibinfo{series}{Egodocuments and History Series}, Vol.~\bibinfo{volume}{52}.
\newblock \bibinfo{publisher}{Paul John Eakin}.
\newblock


\bibitem[\protect\citeauthoryear{Li, Jing, Tong, Yang, He, and Chen}{Li
  et~al\mbox{.}}{2017}]%
        {Li2017}
\bibfield{author}{\bibinfo{person}{Liangyue Li}, \bibinfo{person}{How Jing},
  \bibinfo{person}{Hanghang Tong}, \bibinfo{person}{Jaewon Yang},
  \bibinfo{person}{Qi He}, {and} \bibinfo{person}{Bee-Chung Chen}.}
  \bibinfo{year}{2017}\natexlab{}.
\newblock \showarticletitle{NEMO: Next Career Move Prediction with Contextual
  Embedding}. In \bibinfo{booktitle}{\emph{IW3C2}}. \bibinfo{publisher}{WWW
  2017 Companion}, \bibinfo{address}{Perth}.
\newblock


\bibitem[\protect\citeauthoryear{Negroni}{Negroni}{2005}]%
        {Negroni05}
\bibfield{author}{\bibinfo{person}{Catherine Negroni}.}
  \bibinfo{year}{2005}\natexlab{}.
\newblock \showarticletitle{La reconversion professionnelle volontaire :
  d’une bifurcation professionnelle à une bifurcation biographique}. In
  \bibinfo{booktitle}{\emph{Cahiers internationaux de sociologie}}.
  \bibinfo{publisher}{PUF}, \bibinfo{address}{Perth},
  \bibinfo{pages}{311--331}.
\newblock


\bibitem[\protect\citeauthoryear{Spärck-Jones}{Spärck-Jones}{1999}]%
        {KarenSparck}
\bibfield{author}{\bibinfo{person}{Karen Spärck-Jones}.}
  \bibinfo{year}{1999}\natexlab{}.
\newblock \showarticletitle{Information retrieval and artificial intelligence}.
\newblock \bibinfo{journal}{\emph{Artificial Intelligence}}
  \bibinfo{volume}{114}, \bibinfo{number}{1-2} (\bibinfo{year}{1999}),
  \bibinfo{pages}{257--281}.
\newblock


\bibitem[\protect\citeauthoryear{Thomas}{Thomas}{2013}]%
        {Amosse}
\bibfield{author}{\bibinfo{person}{Amossé Thomas}.}
  \bibinfo{year}{2013}\natexlab{}.
\newblock \showarticletitle{Revisiting the History of Socio-professional
  Classification in France}.
\newblock \bibinfo{journal}{\emph{Annales. Histoire - Sciences Sociales}}
  \bibinfo{volume}{4}, \bibinfo{number}{2} (\bibinfo{year}{2013}).
\newblock


\bibitem[\protect\citeauthoryear{Voorhees and Harman}{Voorhees and
  Harman}{2000}]%
        {Voorhees1999}
\bibfield{author}{\bibinfo{person}{Ellen~M. Voorhees} {and}
  \bibinfo{person}{Donna Harman}.} \bibinfo{year}{2000}\natexlab{}.
\newblock \showarticletitle{Overview of the Eighth Text REtrieval Conference
  (TREC-8)}. \bibinfo{pages}{1--24}.
\newblock


\end{thebibliography}
